# Artificial Intelligence for Neuro MRI Acquisition: A Review

Hongjia Yang[1], Guanhua Wang[2†], Ziyu Li[3], Haoxiang Li[1], Jialan Zheng[1], Yuxin Hu[4], Xiaozhi Cao[4, 5], Congyu Liao[4, 5], Huihui Ye[6], Qiyuan Tian[1*]

[1] School of Biomedical Engineering, Tsinghua University, Beijing, China
[2] University of Michigan, Ann Arbor, MI 48105, USA
[3] Wellcome Centre for Integrative Neuroimaging, FMRIB, Nuffield Department of Clinical Neurosciences, University of Oxford, Oxford, UK
[4] Department of Electrical Engineering, Stanford University, Stanford, California, USA
[5] Department of Radiology, Stanford University, Stanford, California, USA
[6] State Key Laboratory of Extreme Photonics and Instrumentation, College of Optical Science and Engineering, Zhejiang University, Hangzhou, China

[*]Correspondence to: Qiyuan Tian, Ph.D., Center for Biomedical Imaging Research at Tsinghua University, 30 Shuangqing Road, Haidian District, Beijing, China, 100084. E-mail: qiyuantian@tsinghua.edu.cn.

[†]Guanhua Wang is currently at Q Bio, Inc. His contribution was accomplished at the University of Michigan.

Hongjia Yang and Guanhua Wang have contributed equally to this work.

**Abstract.**
Magnetic resonance imaging (MRI) has significantly benefited from the resurgence of artificial intelligence (AI). By leveraging AI's capabilities in large-scale optimization and pattern recognition, innovative methods are transforming the MRI acquisition workflow, including planning, sequence design, and correction of acquisition artifacts. These emerging algorithms demonstrate substantial potential in enhancing the efficiency and throughput of acquisition steps. This review discusses several pivotal AI-based methods in neuro MRI acquisition, focusing on their technological advances, impact on clinical practice, and potential risks.

## 1   Introduction

Magnetic resonance imaging (MRI) is an advanced medical imaging modality that provides comprehensive information on anatomy, microstructure, function, and metabolism. MRI scans are non-invasive, radiation-free, and offer high soft tissue contrast and spatiotemporal resolution, making them widely applicable in clinical practice and scientific research.

Despite its advantages, MRI has a time-consuming and intricate data acquisition process that requires meticulous planning, positioning, and prescription of scans. Variability between operators can lead to inconsistencies in image quality, necessitating standardized and consistent protocols. The scarcity of skilled MRI technologists, particularly in developing regions, further underscores the urgency for automated and intelligent acquisition schemes. Moreover, the extensive scan durations, the risk of peripheral nerve stimulation (PNS), and the specific absorption rate (SAR) constraints limit the utilization of advanced sequences. Complications such as B0 [1] and B1+ [2] field inhomogeneities, along with patient motion, further complicate the acquisition process, potentially resulting in signal degradation and imaging artifacts.

To solve these issues in MRI acquisition, the scientific community has developed numerous innovative solutions over the past four decades, integrating physics, signal processing, and clinical knowledge. The advent of deep learning introduces novel approaches to these challenges by: 1) leveraging extensive historical data repositories, as opposed to the limited datasets typically sourced from the researchers' own institutions or directly collected by the researchers, and 2) employing deep learning's robust capability to model complex, non-linear, and dynamic phenomena. This paper reviews several emerging methodologies, most of which are currently in the proof-of-concept stage yet hold significant promise for practical applications. It also examines recent advancements in data-driven strategies that, while not exclusively reliant on deep neural networks, utilize modern deep learning tools, such as automatic differentiation frameworks like PyTorch [3].

Fig. 1 presents a simplified overview of MRI scanning procedures. In alignment with the other papers of this special issue, this paper delves into key aspects of neuro MRI acquisition: planning, pulse sequence development, and the correction of acquisition-related artifacts. Given the fast-paced advancements in this domain, where new findings are published every week, it is impractical to encompass all relevant studies. This review prioritizes recent studies, emphasizing their technical novelty and the implications for clinical practice, including potential risks.

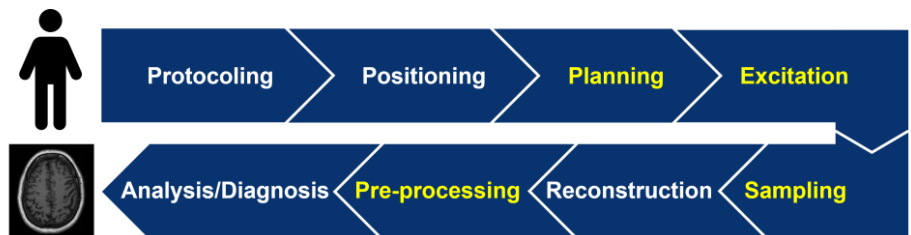

Fig. 1 **A simplified workflow of neuro MRI scans**
Planning, excitation, sampling, and pre-processing parts and their AI applications are reviewed in this paper. Other parts, such as image reconstruction, are discussed in other papers in this special issue.

## 2 AI For Acquisition Planning

Acquisition planning in neuro MRI encompasses many steps, such as patient positioning and shimming, which require extensive operator training and can prolong scan durations. The integration of automated AI technologies offers a promising solution to mitigate these challenges, enhancing data uniformity, diminishing variability introduced by different operators, and potentially shortening scan times.

### 2.1 ROI Selection

Currently, most MRI technicians manually select the region of interest (ROI) based on localizer images. Many data-driven methods that can automatically select the field of view (FOV) have been proposed since the 90s [4-6]. Most methods use classical feature extractors such as scale-invariant feature transform (SIFT) [7]. Modern AI algorithms can significantly enhance this process, using their unparalleled pattern recognition ability. The ROI selection can be formulated as segmentation, detection, or regression tasks, which have been extensively studied by the deep learning community [8,9].

For example, Lei et al. [10] introduced a deep-learning framework for automating FOV prescriptions that uses a 2D convolutional neural network (CNN) to extract features from all image slices. An attention network is then utilized to generate scalars representing the importance of each slice based on the extracted features. Alansary et al. [11] proposed a fully automated method for searching standardized view planes in 3D image acquisitions. This approach adopts a multi-scale reinforcement learning agent framework, employing various Deep Q-network (DQN)-based architectures to solve the reinforcement learning formulation of view planning tasks. The applications of this method in brain and heart MRI showcase real-time target plane detection with high accuracy. Yang et al. [12] introduced a multi-task deep neural network architecture to enhance the learning efficiency and prediction accuracy of brain anatomical landmarks localization. This improvement enables the automatic execution of patient positioning and scan planning during brain MR scanning. (representative images shown in Fig. 2)

AI-based planning methods have been integrated into MRI scanners as products. For instance, GE HealthCare's AIR x™ [13] claims to utilize deep learning to detect patient anatomy automatically and prescribe MRI slices, providing consistent and quantifiable results. Similarly, Canon Medical Systems USA [14] claims to utilize deep learning and machine learning in its Auto Scan Assist to standardize workflow, automating slice alignment for a range of anatomical examinations.

In particular, fetal MRI poses a more challenging problem for localization due to motion. To solve this, Hoffmann et al. [15] developed an integrated and automatic system that utilizes deep neural networks to extract fetal brain and eye positions from rapid full-uterus scout slices. Furthermore, the system deduces the fetal head pose to automatically prescribe slices. The proposed approach reduces the need for repeated acquisitions and FOV adjustments caused by the random orientation of the fetal brain relative to the device axis and fetal motion.

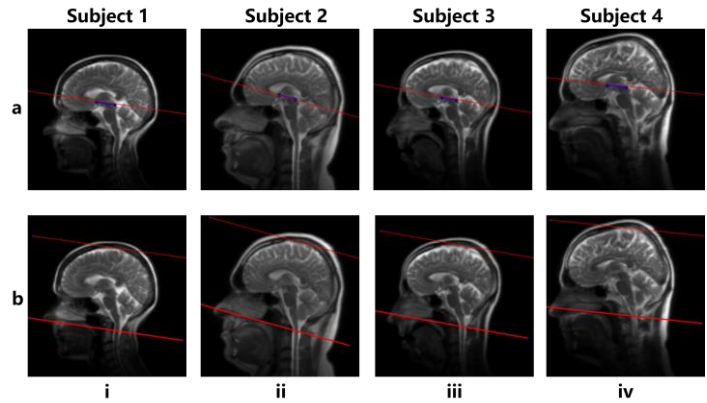

Fig. 2 **Localization of brain anatomical landmarks**
The anterior commissure-posterior commissure (AC-PC) lines (a) and corresponding top/bottom borders (b) automatically determined by a multi-task deep neural network are displayed on sagittal images for four subjects (i-iv). (Image is reproduced from [12] Fig. 7.)

Neural networks excel in localization tasks due to their unparalleled feature extraction capabilities, setting them apart from conventional methodologies that rely on manually designed features [5]. Future research should improve these networks' adaptability to anatomical anomalies [16], a critical step towards harnessing their full potential in wider medical imaging scenarios. Additionally, lightweight neural network models and efficient inference are also crucial, considering the limited computational resources available on most MRI consoles [17].

**2.2 Shimming**

Active shimming typically refers to the adjustment of magnetic main field homogeneity to reduce susceptibility-induced image artifacts and geometric distortions. Modern scanners are usually equipped with first- and second-order shimming [18,19]. As the next-gen shimming technology, local multi-coil shimming [20,21] can further improve B0 homogeneity to mitigate echo-planar imaging (EPI) distortions for structural, functional, and diffusion MRI applications [22-24].

AI algorithms assist in optimizing shimming procedures by rapidly analyzing field maps and suggesting adjustments to achieve optimal homogeneity, especially for high-order shimming.

Becker et al. [25] presented a fast, first-order shimming method in low-field nuclear MR applications, utilizing deep regression with ensembles. The approach leverages input spectra to predict shim currents rapidly, demonstrating a high success rate in improving spectral quality. Its accuracy is expected to be improved by incorporating higher-order shims, localizable information, and special hardware additions. Xu et al. [26] proposed utilizing a combination of automated brain extraction and image-based high-order shimming to compute the shim currents, which was reported to outperform existing rule-based methods. It is worth noting that deep learning is solely used for brain extraction in this approach and has not been applied in the shimming process.

For addressing time-varying spatially non-linear B0 inhomogeneity, Zhang et al. [27] developed an auto-differentiation framework that optimizes radio frequency (RF) pulses and time-varying high-order spatially non-linear ΔB0 fields to achieve refined magnetization profiles within fixed voxel intervals, enabling restricted slice excitation and refocusing. The proposed framework significantly enhances the fidelity of restricted-slice excitation and refocusing for fetal MRI. This advancement can facilitate applications such as zoomed fast spin-echo (FSE) MRI and other related areas [28].

**2.3 Protocol selection**

For automatic planning, in addition to ROI selection and shimming, future studies may further extend the idea of automatic prescription. Choosing the right protocol is crucial in MRI planning. Radiologists and MRI technicians need to consider various factors, including the patient's condition and the specific pathology being investigated. They also need to adjust protocols in real-time based on the latest acquisitions. When issues like motion artifacts occur, it is important to revise and repeat sequences. This makes protocol selection a complex task that involves recognizing patterns, evaluating images, and making decisions. Modern neural networks greatly assist in assessing and evaluating image quality [29]. Additionally, reinforcement learning plays a key role in choosing the appropriate actions [30]. Currently, most studies focus on offline protocol prescriptions based on text or feature (i.e., indications) information [31-33]. Future works should focus on real-time protocol design, using both text and image information. Multi-modal models, especially the recent visual-language models (VLMs), may play a central role in this task [34].

# 3 AI for Sequence Design

MRI acquisition strategy is highly programmable and enables flexibility in contrast mechanisms and encoding strategies. Innovative sequence designs have enabled many novel imaging applications since the birth of MRI.

MRI sequence design encompasses a multitude of parameters. Parameters like repetition time (TR), echo time (TE), inversion time (TI), and flip angle (FA) determine the signal-to-noise ratio (SNR) and image contrast. Beyond these basic parameters, RF pulse design can optimize the waveforms to achieve superior excitation profiles while maintaining robustness to B0/B1+ inhomogeneity. A pulse sequence can also be tailored to desired image contrast, robustness to motion, and improve quantification accuracy [35,36]. Another impactful outcome of sequence optimization is to improve patient comfort, such as reducing SAR and PNS. Generally speaking, spectral and spatially selective RF pulses are two critical topics in advanced sequence design that contribute to applications such as inner-volume excitation and spectral imaging [37-39].

The optimization process itself presents a significant challenge. MRI sequences often involve enormous degrees of freedom, especially considering advanced acquisition strategies, making optimization a complex, high-dimensional problem. AI algorithms, particularly machine learning, excel in handling large-scale optimization problems. They can efficiently search through parameter spaces, finding the optimal combination for desired outcomes.

### 3.1 General RF pulse design

For RF pulse design, multiple methods approach it as an inverse problem where the Bloch equation is posed as the forward model. Gradient methods are employed to invert the problem [36,40,41]. In contrast, Vinding et al. [42] trained a CNN to invert the forward Bloch equation directly. Luo et al. [40] optimized a spatially selective sampling pattern using Limited-memory Broyden-Fletcher-Goldfarb-Shanno (L-BFGS). Loktyushin et al. [41] proposed a supervised learning and fully differentiable framework to automatically generate MRI sequences based on target contrast of interest. Due to the non-convexity of RF pulse optimization problems, these methods are often troubled by local minima. Efficient parameterization strategies such as B-splines [40,43] may help alleviate this problem.

In addition to gradient methods, Zhu et al. [44] optimized pulse sequence development using a Bayesian derivative of reinforcement learning. Despite its high flexibility, this method necessitates a prolonged search process to obtain the optimal result. Similarly, Shin et al. [45] used reinforcement learning to generate initial guesses of general-purpose excitation and B1+ insensitive pulses, then refined these results using gradient methods to reduce the non-convexity. Black-box algorithms can also be applied. For example, Hoinkiss et al. [36] proposed a domain-specific language (DSL) to formulate the corresponding machine-learning problem to optimize MRI sequences. In addition to improving SNR and image contrast, this approach also addresses constrained parameters, enhancing robustness against subject motion and geometric distortion. Black-box algorithms and reinforcement learning may explore a broader solver space than gradient methods, but they suffer from low sampling efficiency. Similar to the method proposed by Shin et al. [45], combining them with gradient methods may lead to faster convergence.

In particular, as a quantitative imaging sequence, the design for MR fingerprinting sequences [46] receives much attention, involving a magnitude of parameters and poses a large-scale optimization problem. Several works [47-49] applied gradient-based methods using techniques such as auto-differentiation and parameterization, while Jordan et al. [50] used genetic algorithms (representative images shown in Fig. 3(a)).

**3.2 K-space sampling trajectory design**

MRI scans are typically lengthy, especially for developing and aging subjects and certain populations of patients, and the scan time is proportional to the amount of acquired data. Deep learning has also presented novel opportunities for k-space trajectory design, further improving manual, rule-based designs. Gradient waveforms can be intelligently optimized to accelerate image acquisition while maintaining or even enhancing image quality. Optimization may address side effects like eddy currents, acoustic noise, and mechanical vibrations.

Generally, sampling patterns can be categorized into Cartesian and non-Cartesian sampling patterns, while intersections such as wave-encoding and stack-of-stars are also widely used [51,52].

The Cartesian sampling pattern usually presents itself as a collection of discrete locations. A salient metric in assessing the efficacy of a sampling pattern lies in its image reconstruction results. Consequently, the majority of these techniques are split into two stages: the design of the sampling pattern and the subsequent image reconstruction. For instance, Sherry et al. [53] approached this challenge as bilevel optimization, where the upper level focuses on the sampling pattern, while the lower level addresses regularized MRI reconstructions. This bilevel problem is tackled using the Limited-memory Broyden-Fletcher-Goldfarb-Shanno with Bounds (L-BFGS-B) algorithm [54]. Zibetti et al. [55] and Gozcu et al. [56] used greedy algorithms to search for optimal sampling patterns. Haldar et al. [57] and Seeger et al. [58] used Bayesian methods to optimize Cartesian sampling patterns. To enable gradient methods for discrete variables, Huijben et al. [59] used reparameterization strategies.

Bahadir et al. [60] proposed an end-to-end deep learning method, LOUPE (Learning-based Optimization of the Under-sampling Pattern), which includes two primary components: a network dedicated to optimizing the sampling pattern, which discerns a probabilistic k-space mask and a secondary element for reconstruction using CNNs. LOUPE's results underscore its ability to generate anatomy-centric masks that outperform conventional methodologies like uniform random or variable density sampling.

In contrast to Cartesian sampling, non-Cartesian sampling patterns are continuous variables, thus amenable to gradient methods. The PILOT (Physics Informed Learned Optimal Trajectory) [61] technique adopts a dual-phase approach, with a distinctive emphasis on incorporating physical constraints, such as gradient and slew rate limits, to ensure the practicality of trajectories. Additionally, PILOT is versatile and capable of adapting trajectories to suit tasks beyond reconstruction, including segmentation. Weiss et al. [61] and Alush-Aben et al. [62] applied auto-differentiation to non-uniform fast Fourier transform (NUFFT) to calculate the Jacobian required in optimization, while Wang et al. [43,63] analytically derived the Jacobian operators used in the gradient methods (representative images shown in Fig. 3(b)), for which Gossard et al. [64] did more analysis into the non-convexity of this problem. As an extension, Wang et al. [65] further improved the PNS behavior in addition to image quality for 3D sampling trajectories. Non-Cartesian sampling trajectories are bounded by maximum gradient strengths and slew rates. Weiss et al. [61] proposed the use of a soft penalty, while Radhakrishna et al. [66] adopted a projection gradient descent approach.

A critical topic for sampling pattern design is active sampling patterns, which predict the next sampling locations on the fly. Pineda et al. [67] used reinforcement learning to find optimal subsets, with image quality as a reward, while Wang et al. [68] and Sanchez et al. [69] used a Bayesian approach.

Even though RF pulses and receiving k-space sampling are discussed separately here, they are part of a coupled optimization process, especially for complex sequences such as fast spin-echo. Dang et al. [35] and Hoinkiss et al. [36] used systemic point-spread-function (PSF), considering relaxation, frequency sampling, and reconstruction as an optimization criterion, which is preferable to optimizing each component individually.

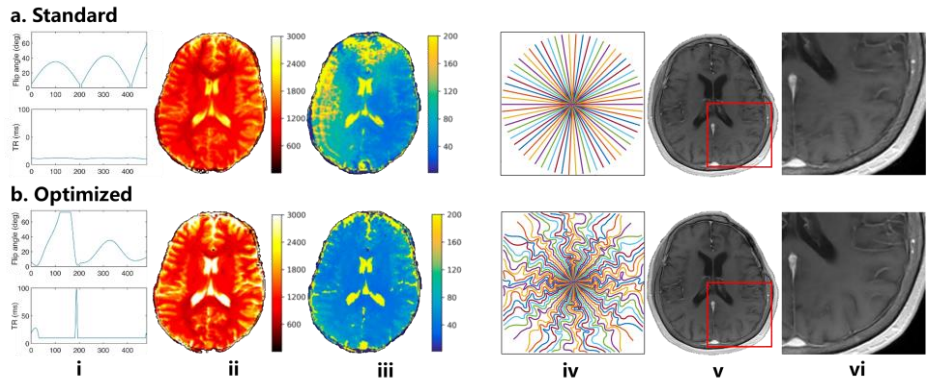

Fig. 3 **Comparison of standard and AI-optimized sequence designs**
Comparison of standard images (a, i-iii) and sequences/k-space trajectories (a, iv-vi) and optimized images (b, i-iii) and sequences/k-space trajectories (b, iv-vi) in sequence design. (i-iii) In vivo results for T1 (red) and T2 (blue) map simulations of a standard human-designed sequence and an optimized sequence incorporating phase variation. The error is modeled as a time-independent phase that varies quadratically along a chosen direction. Experimentally, one finds that this direction varies randomly from one scan to the next. (iv-vi) Examples from the simulation experiment using the unrolled neural network (UNN)-based reconstruction algorithm, with 32 shots for T1w contrast. Red boxes indicate the zoom-in region. (Image is reproduced from [50] Fig. 2, 3, [43] Fig. 2, 11.)

In summary, these studies confirm that MR pulse sequences can be tailored using optimization methods, demonstrating superior outcomes compared to traditional, rule-based approaches. Notably, as some studies have shown, the learned protocols can develop anatomy-specific trajectories [43,60-62]. This feature may serve as a double-edged sword, enhancing performance for certain anatomies while potentially compromising others. However, other works demonstrate that learned sampling trajectories generalize well across different field strengths, vendors, and contrasts [43,70]. Generally speaking, the learned sequences still sample the key k-space information but with an optimized strategy that may emphasize certain details (especially the low-frequency components) and exploit relationships such as conjugate symmetry and multi-coil covariance. Future research should further explore the generalizability of these optimized sequences.

## 4    AI for Acquisition Artifacts Correction

MRI is prone to artifacts during acquisition due to field imperfections (e.g., B0 inhomogeneity, B1+ inhomogeneity, and eddy currents). The artifacts substantially degrade the image quality and can be a confounding factor for accurate diagnosis and image analysis. AI techniques have demonstrated improved performance in correcting artifacts and enhancing image quality.

**4.1 B0 off-resonance**

Main field inhomogeneity is a major source of image artifacts. It leads to a non-uniform precession frequency of the spins, which is known as the "off-resonance" effect, and causes several acquisition artifacts such as geometric distortions, signal loss, and blurring [71]. B0 inhomogeneity results from the intrinsic non-uniform magnetic susceptibility of the scanned subject and the main magnet's imperfections. It is particularly strong at the air-tissue interface and/or at a higher B0 field strength (e.g., 7T) [1]. Other factors, such as chemical shift [72] and metal implants [73] also contribute to B0 inhomogeneity. The B0 field inhomogeneity introduces geometric distortions to EPI which affects numerous diffusion and functional MRI applications where EPI has been the workhorse [70,74]. It also causes blurring in spiral imaging which is adopted in cardiac imaging [75], diffusion [70], and functional MRI [76]. Applications prone to susceptibility variations also suffer from artifacts brought by B0 field inhomogeneity. For example, the imaging of patients of certain populations can be complicated by metal implants such as surgical clips, dental fillings, fixation screws, surgical pins, or intra-cortical electrodes [73].

Conventionally, the B0 field inhomogeneity can be measured using a multi-echo GRE field mapping sequence [77] or a pair of EPI acquired along opposite phase encoding directions, coupled with image processing methods such as the widely adopted "topup" function from the FSL [78,79]. Recent developments have enabled a more integrated way of correcting B0 field inhomogeneity which incorporates the blip-reversed

EPI acquisition into the imaging framework to allow for correction during reconstruction.

Deep learning is leveraged for B0 field inhomogeneity correction mainly in three ways: 1) B0 field inhomogeneity-induced artifacts correction using supervised learning; 2) B0 field estimation using EPI sequences along opposite phase encoding directions using self-supervised learning; 3) B0 field estimation assisted by deep learning-based image synthesis.

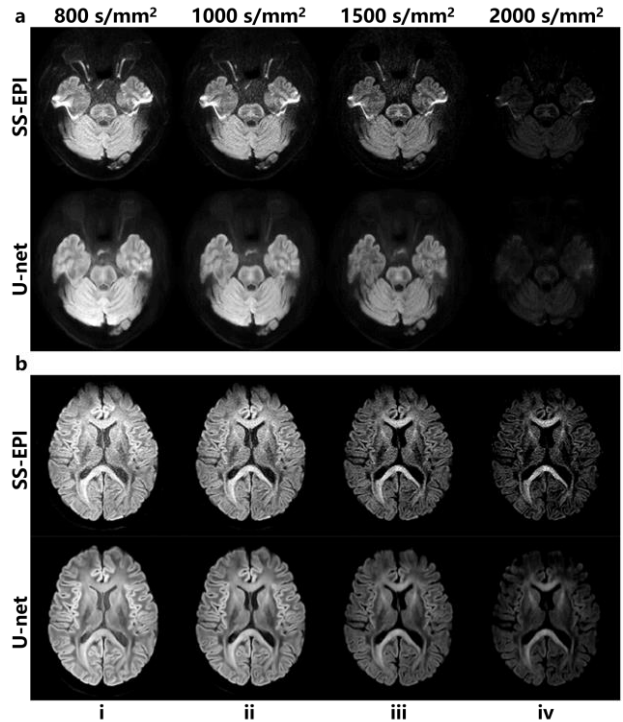

Fig. 4 **B0-related artifacts correction**
Distortion correction results of SS-EPI diffusion-weighted images (DWIs) with different b-values using U-net. The images with b-values = 800, 1500, and 2000 s/mm$^2$ were simulated from the b-value = 1000 s/mm$^2$ images using the monoexponential signal decay model. Two slices are shown in a and b. (Image is reproduced from [80] Fig. 9.)

For the supervised learning-based approaches, Hu et al. [80] proposed to leverage supervised learning with a 2D U-Net to correct for B0 inhomogeneity-induced image distortion in single-shot EPI (SS-EPI) (representative images shown in Fig. 4). The point-spread-function encoded EPI (PSF-EPI) [81] is used as the distortion-free target. The distortion correction performance is significantly superior to those from "topup" and field mapping-based methods thanks to the superior anatomical fidelity provided by PSF-EPI. Building upon this work, Ye et al. [82] further enabled simultaneous distortion correction and super-resolution of low-resolution SS-EPI using a generative adversarial network (GAN) [83] with high-resolution PSF-EPI as targets. Haskell et al. [84] developed "FieldMapNet" to map each image in a spiral-in blood oxygen level-dependent (BOLD) functional MRI acquisition to a corresponding B0 field map and perform customized B0 correction at each functional MRI timepoint, which reduces the distortion artifacts and blurring.

The presence of metal can result in strong susceptibility variations, leading to B0 off-resonance [73]. Yan et al. [85] leveraged a GAN to correct artifacts in functional MRI caused by metal implants-induced B0 inhomogeneity. They simulated the signal loss in BOLD images in regions mostly affected by metal implants to obtain training data and demonstrated their model's capacity to recover the lost BOLD signal in both healthy subjects and patients.

While supervised deep learning methods exhibit satisfactory performance in correcting image distortion (Fig. 3), the difficulty in obtaining sufficient reference data for training could be an obstacle to their wider adoption. For example, the acquisition of distortion-free PSF-EPI [80] requires more than 20-fold scan time per volume compared to SS-EPI with matched resolution. Data simulation and augmentation are useful methods for addressing training data requirements for supervised deep learning [84,85].

The second group of work aims to achieve self-supervised "topup"-like correction with improved accuracy and computational efficiency without requiring distortion-free reference. Zahneisen et al. [86] developed "flow-net" which estimated the B0 field map from reversed phase-encoding EPI pair with shortened processing time compared to "topup". The network is trained in an unsupervised manner by minimizing the difference between blip-up and blip-down images after applying the estimated field map, similar to "topup". In the meantime, S-Net [87] with different architecture was proposed following similar principles. The performance of flow-Net and S-Net are improved for diffusion MRI data distortion correction by incorporating "topup" and "eddy" [88] into the distortion correction framework and leveraging fiber orientation distribution derived from the diffusion data [89]. More recently, FD-Net [90] was proposed, which applies the forward distortion model to the distortion-corrected image along both phase-encoding directions and enforces the similarity between the forward-distorted and the input images.

These methods demonstrate improved performance and processing speed compared to "topup", but still require reversed phase-encoding EPI pair, which is not often available in practice.

Image synthesis methods from the third group leverage complementary information from other modalities routinely acquired to assist the distortion correction when a reversed phase-encoded EPI pair is unavailable. Synb0-DisCo [91] and Synb0 [92] proposed to synthesize a distortion-free b=0 image from a T1-weighted image, which enables the prediction of field map using "topup" with the synthetic b=0 and the acquired b=0 along one phase-encoding direction as inputs. In contrast, DeepAnat [93] generated a distorted T1-weighted image from the distorted diffusion data for improved non-linear co-registration to the distortion-free T1-weighted image space. The resultant warp field is then used for correcting the geometric distortions in the diffusion data. Although these supervised image synthesis models also require training data, abundant multi-contrast training data are available in large-scale public datasets (e.g., Human Connectome Project [94], UK Biobank [95]).

It is worth noting the possibility of integrating the distortion correction into a deep learning-based reconstruction framework. Shan et al. [96] demonstrated incorporating the distortion operator into the deep learning model-based reconstruction provides improved accuracy and processing speed for distortion correction from gradient nonlinearity. It is expected to be feasible to apply such methods in addressing the distortions from B0 inhomogeneity.

**4.2 B1+ field-related artifacts**

Another source of image artifacts is the inhomogeneity in the magnetic transmit field (B1+), which is especially severe at higher field strengths (e.g., 7T) due to more significant wave interference effects. B1+ inhomogeneity leads to substantial image artifacts such as image contrast alteration and signal dropout for a wide range of high-field applications such as structural [97], functional [98], and diffusion [99] MRI for both head [100] and body [101] imaging.

Recently, there has been growing interest in reducing B1+ inhomogeneity for highfield imaging using multiple transmission channels, known as parallel transmission (pTx) [97]. This technique utilizes specially designed pulses that incorporate measured B1+ field in multiple transmission channels to achieve higher uniformity of the combined field, which requires additional scan time for mapping the B1+ and $\Delta$B0 field. Furthermore, time-consuming non-convex optimization problems are often involved in designing such pulses [102]. The Universal Pulse method that pre-calculates a universal pTx pulse waveform using B1+ and $\Delta$B0 field maps acquired from a representative sample of the adult population has been developed to overcome these hurdles [103].

Machine and deep learning methods are utilized for more efficient pTx workflow. For example, SmartPulse [104] was developed to address the inaccuracies from subject variability when a single universal pulse is used. It selects a universal pulse from a group of universal pulses optimized on different groups of subjects based on the features of the scanned subject. Ianni et al. [105] leveraged machine learning for individual RF shim predictions based on subjects' shape-related features, improving the scan and computational efficiency. More recently, deep learning-based methods are also developed for efficiently estimating B1+ field distribution and mitigating B1+ inhomogeneity-induced artifacts. Plumley et al. [106] tackled head motion-induced B1+ field inaccuracy for pTx using motion detection and a conditional GAN. Krueger et al. [107]

predicted the B1+ field from localizers using a 2D U-Net. Ma et al. [108] proposed another way to reduce B1+-induced artifacts by directly predicting artifact-suppressed pTx images from a conventionally acquired single transmission image using a 2D U-Net (representative images shown in Fig. 5).

Additionally, the inhomogeneity in the receiver coil (B1-) will lead to biased image intensity. Bias field correction is routinely performed in image processing pipelines using software such as SPM [109] and FSL [79]. Deep learning-based image enhancement offers a promising avenue to improve the image quality and accelerate the processing for such correction. Chuang et al. [110] utilized a 3D GAN to simultaneously correct for bias field and perform accurate brain extraction for EPI, which exhibits improved performance compared to conventional methods. Harrevelt et al. [111] demonstrated the efficacy of deep learning-based bias field correction for T2w prostate imaging at 7T.

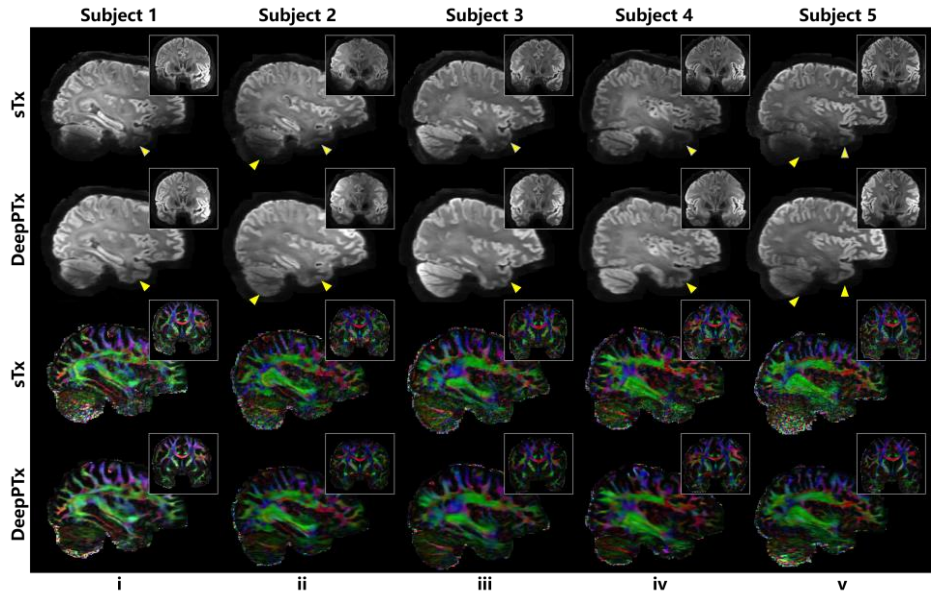

Fig. 5 **B1-related artifacts correction**
Testing of proposed deep learning model on unseen data of 5 new subjects randomly chosen from the 7 T Human Connectome Project database. For each case, a representative sagittal slice from mean diffusion-weighted images at b=1000 s/mm2 (first and second rows) and color-coded fractional anisotropy maps (third and fourth rows) are shown, with a representative coronal slice shown in the insert. The final model with tuned hyperparameters was trained on the entire dataset of 5 subjects. The use of the learned model substantially enhanced the image quality by effectively restoring the signal dropout observed in the lower brain regions (as marked by yellow arrowheads), producing color-coded fractional anisotropy maps that presented substantially reduced noise levels in those challenging regions. (Image is reproduced from [108] Figure 9. )

# 5    Discussion

Our review focused on AI applications in MRI acquisition. Compared to traditional MRI acquisition methods, AI-based approaches provide several advantages, including reduced MRI acquisition time, decreased manual workload, efficient parameter searches and optimizations, and a notable enhancement in image quality.

However, the limitations introduced by data-driven methods must be recognized and addressed before clinical implementation. The robustness and generalization capabilities of AI-driven MRI acquisition techniques require further scrutiny. For instance, models trained solely on healthy cohorts may underperform or pose risks when encountering outlier cases, such as unfamiliar pathologies or severe field distortions. Moreover, the specificity of existing models to certain body regions calls for research into the feasibility of developing versatile systems applicable across various anatomical areas. Despite data scarcity and imbalance, future methodologies must ensure accurate planning across different MRI machines, imaging sequences, contrasts, patient orientations, and demographics [112,113].

Addressing these challenges necessitates a collaborative endeavor within the research community. Developing and validating robust algorithms will benefit from comprehensive, pathological, multicenter datasets, complemented by reader studies. The creation of accessible, physics-based simulation tools to mimic outliers, such as motion and strong field distortions, will also be vital. Methodologically, it is crucial to focus on algorithms that inherently resist domain drift, emphasizing efficient fine-tuning strategies and developing models capable of few-shot learning. The exploration of foundational and large-scale models and the inclusion of imaging physics in AI workflows can also enhance robustness [47,63,114], despite the current limitations of physics models due to oversimplified physics assumptions [115,116].

AI-based MRI planning methods must be designed with practical use scenarios in mind, efficiently generating planning results from pre-scan images to ensure consistent subject positioning between pre-scan and subsequent imaging sequences, which is especially critical for shimming [117].

In sequence design, optimized sequences are less sensitive to new image-domain indications, partially because these sequences are optimized in the frequency domain. Most machine learning-optimized sequences have not accounted for significant non-idealities, such as severe B0 inhomogeneity, which warrants further investigation. Future studies should aim to comprehensively optimize all sequence components—RF pulses, gradient waveforms, and timing—while also considering factors like tissue parameters and partial volume effects. This end-to-end optimization approach presents a large-scale non-convex optimization challenge, highlighting the importance of more accurate Bloch simulators that consider factors such as diffusion, chemical exchange, and magnetization transfer, albeit at a greater computational cost [47,63,114].

Major acquisition artifacts, particularly those induced by B0 and B1+ field inhomogeneities, are significant challenges. For B0-induced distortions, recent studies have replaced traditional methods with neural networks to accelerate processing [78,79]. This approach allows for rapid inference post-training after complex network optimization during the training phase. Inspired by the methods in [118-123], a promising future direction is the accurate and fast joint reconstruction of B0 and distortion-corrected images, potentially facilitated by leveraging unrolled networks in model-based reconstruction strategies. Another approach for distortion correction leverages auxiliary contrasts and CNN-based image synthesis to estimate the B0 field map, even in the absence of paired EPI images with opposite phase-encoding directions [86-90]. These methods primarily utilize CNNs' exceptional image generation capabilities to facilitate nonlinear registration by harmonizing image contrast and resolution. For B1+ correction, despite the efficacy of pTx, its practical application is hampered by the need for additional scans for B0 and B1+ mapping and complex non-convex optimizations. Machine learning can assist at various stages of the pTx workflow, including pulse design [104], shimming [105], and accounting for motion [106].

Future efforts should also concentrate on validating and integrating AI-enhanced solutions into routine clinical practice to fully realize their potential in MRI acquisition and improve patient care. The real-time integration of AI-based strategy planners into MRI scanner workflows requires substantial engineering efforts. Several vendors and research groups are prototyping "one-click" MRI solutions [124], which could significantly improve the reproducibility of functional MRI, diffusion MRI, and other neuroimaging applications that demand experienced operators. Addressing engineering challenges, such as ensuring efficient and rapid communication between AI-based MRI acquisition algorithms and MRI scanner consoles, is crucial for deploying real-time algorithms [125]. The availability of high-speed networking solutions, such as fiber Ethernet, could mitigate data transfer bottlenecks. Additionally, more open access to raw data and control application programming interfaces (APIs) by vendors, along with the integrated deployment of algorithms, would greatly support the development of AI-driven methods. Community software tools, such as ODIN [126], Pulseq [127], and TOPPE [128] for sequence programming, alongside BART [129] and Gadgetron [130] for reconstruction, play a pivotal role in facilitating the development of AI-driven methodologies.

# 6 Conclusion

AI-based MRI acquisition is an active research field. With the emergence of robust AI algorithms and models, accompanied by advancements in technologies guiding the design of MRI acquisition, we can foresee the development of more efficient MRI acquisition techniques. These progressions are positioned to significantly improve the application of MRI in both clinical diagnosis and scientific research.